\documentclass[aps,twocolumn,nofootinbib,showkeys,notitlepage,superscriptaddress]{revtex4}

\usepackage{graphicx}
\usepackage{dcolumn}
\usepackage{bm}
\usepackage{amsmath}
\usepackage{float}
\usepackage{multirow}
\usepackage{slashed}
\usepackage{xcolor}
\usepackage{physics}
\usepackage{multirow}
\usepackage[colorlinks=true, pdfstartview=FitV, bookmarks=true, bookmarksnumbered=true, breaklinks]{hyperref}
\usepackage{mathtools,braket}
\usepackage{soul}
\usepackage{lipsum}  
\usepackage{color}
\definecolor{blue}{rgb}{0.0, 0.0, 1.0}
\definecolor{red}{rgb}{1.0, 0.0, 0.0}
\definecolor{royalblue}{rgb}{0.0, 0.14, 0.4}
\hypersetup{linkcolor=royalblue, citecolor=blue, urlcolor=royalblue}
\newcommand{\mytitle}[1]{\medskip {\em #1.---}}
\def\orcid#1{\kern .08em\href{https://orcid.org/#1}{\includegraphics[keepaspectratio,width=0.7em]{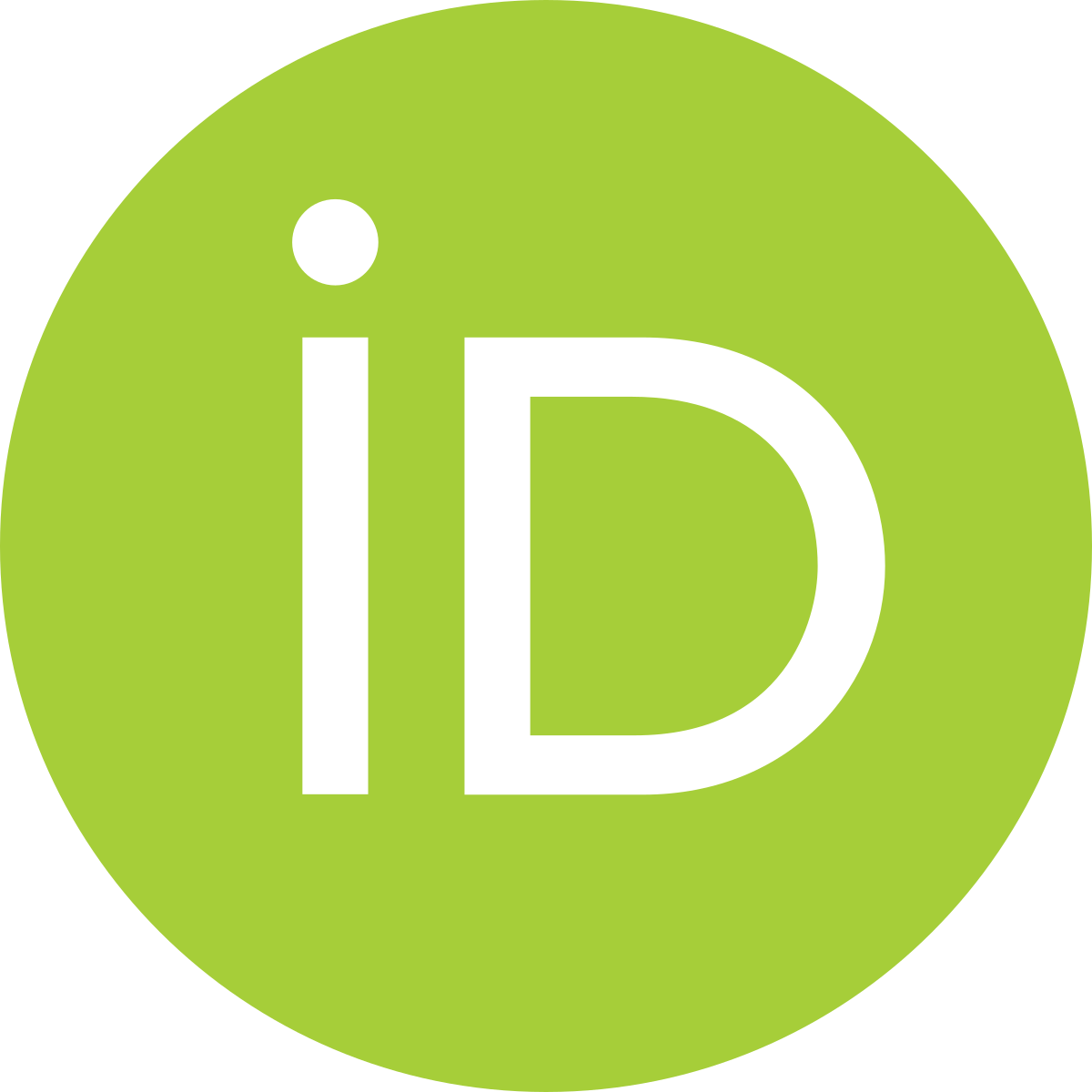}}}

\usepackage{hyperref}
\hypersetup{colorlinks=true,citecolor=blue,linkcolor=blue,urlcolor=blue}

\usepackage[mathlines]{lineno}

\begin{document}

\title{ 
Independence of current components, polarization vectors, and reference frames \\
in the light-front quark model analysis of meson decay constants}

\author{Ahmad Jafar Arifi\orcid{0000-0002-9530-8993}}
\email{ahmad.jafar.arifi@apctp.org}
\affiliation{Asia Pacific Center for Theoretical Physics, Pohang, Gyeongbuk 37673, Korea}

\author{Ho-Meoyng Choi\orcid{0000-0003-1604-7279}}
\email{homyoung@knu.ac.kr}
\affiliation{Department of Physics Education, Teachers College, Kyungpook National University, Daegu 41566, Korea}

\author{Chueng-Ryong Ji\orcid{0000-0002-3024-5186}}
\email{ji@ncsu.edu}
\affiliation{Department of Physics, North Carolina State University, Raleigh, NC 27695-8202, USA}

\author{Yongseok Oh\orcid{0000-0001-9822-8975}}
\email{yohphy@knu.ac.kr}
\affiliation{Department of Physics, Kyungpook National University, Daegu 41566, Korea}
\affiliation{Asia Pacific Center for Theoretical Physics, Pohang, Gyeongbuk 37673, Korea}

\date{\today}

\begin{abstract}

The issue of resulting in the same physical observables with different current components, 
in particular from the minus current, has been challenging in the light-front quark model (LFQM) even 
for the computation of the two-point functions such as meson decay constants.
At the level of one-body current matrix element computation, we show the uniqueness of pseudoscalar 
and vector meson decay constants using all available components including the minus component of 
the current in the LFQM consistent with the Bakamjian-Thomas construction. 
Regardless of the current components, the polarization vectors, and the reference frames, the meson decay 
constants are uniquely determined in the non-interacting constituent quark and antiquark basis while the 
interactions of the constituents are added to the meson mass operator in the LFQM.

\end{abstract}

\maketitle 


\mytitle{Introduction}
Light-front dynamics (LFD)~\cite{Terentev76,LB80,BPP97} is a useful framework for studying hadron structures 
with its direct applications in Minkowski space.
The distinct features of LFD compared to other forms of Hamiltonian dynamics include that the rational energy-momentum 
dispersion relation in the LFD induces the suppression of vacuum fluctuations and that the LFD carries the maximal number 
(seven) of the kinematic generators of transformations for the Poincar\'e group.

The light-front quark model (LFQM) based on the LFD has been quite successful in describing the mass spectra 
and electroweak properties of mesons by treating mesons as quark-antiquark bound 
states~\cite{Jaus90,Jaus91,CCH96,CP05,CJ97,CJ99a,CJ07,CJLR15,ACJO22,DF97,DFPS03,BCJ02,LMV17,CLLSY18}.
Typically in the LFQM~\cite{Jaus90,Jaus91,CCH96,CP05,CJ97,CJ99a,CJ07,CJLR15,ACJO22}, the constituent quark ($Q$) and 
antiquark ($\bar{Q}$) are constrained to be on their respective mass shells, and the spin-orbit (SO) wave function is thus obtained by 
the interaction-independent Melosh transformation~\cite{Melosh74} from the ordinary equal-time static representation.
While the hadronic form factors and decay constants are obtained from the matrix elements of a one-body current 
directly in the three-dimensional light front (LF) momentum space effectively with the plus current, $J^+=J^0 + J^3$, 
the calculation with different current components such as the transverse current 
${\bf J}_\perp$ and the minus current, 
$J^-=J^0-J^3$, should in principle yield the same results of the hadronic form factors and decay constants as the physical 
observables must be Lorentz invariant. 
However, in practice, the issue of resulting in the same physical observables with different current components has been 
challenging in LFQM and led discussions on the Fock space truncation~\cite{LMV17b}, the zero-mode 
contribution~\cite{Jaus99,CCH03}, etc., in a variety of contexts~\cite{CDKM98,CCKP88,Keister94,SJC95,FFS93,LLCLV21}. 
Thus, clarifying this long-standing issue even in the two-point function level, such as the computation of decay constants, 
is of great importance to construct a reliable light-front model to study hadron structure.

Focusing on the vector meson decay amplitudes with the matrix element of one-body current~\cite{CJ13}, 
two of us showed that the decay constants obtained from $J^+$ with longitudinal polarization and ${\bf J}_\perp$
with transverse polarization  are numerically the same 
by imposing the on-shellness of the constituents consistently throughout the LFQM analysis. In fact, it was demonstrated that those two decay constants obtained from 
using the so-called ``Type II"~\cite{CJ13} link between the manifestly
covariant Bethe-Salpeter (BS) model and the standard LFQM are exactly equal to those obtained directly in the standard LFQM imposing the on-shellness of the constituents.

This on-mass shell condition is equivalent to imposing the four-momentum conservation $P=p_1+p_2$ at the meson-quark vertex, 
where $P$ and $p_{1(2)}$ are the meson and quark (antiquark) momenta, respectively, which implies the self-consistent replacement of the physical 
meson mass $M$ with the invariant mass $M_0$ of the quark-antiquark system. 
The generalization of the results in Ref.~\cite{CJ13} to any possible combination of
current components and of polarization is the main object of the present work.

We notice in retrospect that this condition for the one-body current matrix element computation is consistent with the 
Bakamjian-Thomas (BT) construction~\cite{BT53,KP91} up to that level of computation, where the meson state is 
constructed by the noninteracting $Q\bar{Q}$ representations while the interaction is included into the mass 
operator $M:= M_0 + V_{Q\bar{Q}}$ to satisfy the group structure or commutation relations. 
The main purpose of the present work is to demonstrate that the long-standing issue of resulting in the same 
physical observables with different current components can be resolved for the two-point physical observables, 
explicitly in the analysis of the decay constants for the one-body current matrix element computation 
with the aforementioned self-consistent condition stemmed from the BT construction. 
We note that the meson system of the constituent quark and antiquark presented in this work is immune 
to the limitation of the BT construction regarding the cluster separability for the systems of more than 
two particles~\cite{Keister-Polyzou-2012}.

Within the scope described above, we show for the first time the uniqueness of pseudoscalar and vector meson 
decay constants using all available components of the current in our LFQM being consistent with the BT construction 
for the one-body current matrix element computation. 
We explicitly demonstrate that the same decay constants are resulted not only for all possible current components  
but also for the polarization vectors independent of the reference frame.
Our explicit demonstration is in fact related to the Lorentz invariant property that could not be obtained in the relativistic 
quark models based on LFD without implementing the aforementioned self-consistency condition.

\mytitle{Theoretical framework}
While our demonstration can be applied to the mesons composed of unequal-mass constituents in general,
here we focus on the equal mass case of the constituents for simplicity.
The essential aspect of the standard LFQM for the meson state~\cite{Jaus91,Jaus90,CCH96,CP05,CJ97,CJ99a,CJ07} 
is to saturate the Fock state expansion by the constituent quark and antiquark and treat the Fock state 
in a noninteracting representation.
The interactions are then encoded in the LF wave function $\Psi_{\lambda_1\lambda_2}^{JJ_z}({\bf p}_1,{\bf p}_2)$, 
which is the mass eigenfunction.
The meson state $\ket{M (P,J, J_z)} \equiv \ket{\cal M}$ of momentum $P$ and spin state $(J,J_z)$ can be constructed as
\begin{eqnarray}\label{eq:1}
\ket{\mathcal{M}} 
&=& \int \left[ {\rm d}^3{\bf p}_1 \right] \left[ {\rm d}^3{\bf p}_2 \right]  2(2\pi)^3 \delta^3 \left({\bf P}-{\bf p}_1-{\bf p}_2 \right) 
\nonumber\\ && \times \mbox{} 
\sum_{\lambda_1,\lambda_2} \Psi_{\lambda_1 \lambda_2}^{JJ_z}({\bf p}_1,{\bf p}_2) 
\ket{Q(p_1,\lambda_1) \bar{Q}(p_2,\lambda_2) },
\quad
\end{eqnarray}
where $p^\mu_i$ and $\lambda_i$ are the momenta and the helicities of the on-mass shell $(p_i^2=m^2_i)$ constituent quarks, 
respectively.
For the equal mass case, we set $m_i = m$.
Here, ${\bf p}=(p^+,{\bf p}_\perp)$ and $\left[ {\rm d}^3{\bf p}_i \right] \equiv {\rm d}p_i^+ {\rm d}^2\mathbf{p}_{i\perp}/(16\pi^3)$.
The LF relative momentum variables $(x, {\bf k}_\perp)$ are defined as $x_i=p^+_i/P^+$ and  
${\bf k}_{i\perp} = {\bf p}_{i\perp} - x_i {\bf P}_\perp$, which satisfy $\sum_i x_i=1$ and $\sum_i {\bf k}_{i\perp}=0$.
By setting $x \equiv x_1$ and ${\bf k}_{\perp}\equiv{\bf k}_{1\perp}$, we decompose the LF wave function as
$\Psi^{JJ_z}_{\lambda_1\lambda_2} (x, \mathbf{k}_{\bot}) = \phi(x, \mathbf{k}_\bot) 
\mathcal{R}^{JJ_z}_{\lambda_1\lambda_2}(x, \mathbf{k}_\bot)$, where $\phi(x, \mathbf{k}_\bot)$ is the radial wave function and 
$\mathcal{R}^{JJ_z}_{\lambda_1\lambda_2 }$ is the SO wave function obtained by the interaction-independent 
Melosh transformation.

The covariant forms of the SO wave functions are $\mathcal{R}^{JJ_z}_{\lambda_1\lambda_2} = \bar{u}_{\lambda_1}^{}(p_1)  
\Gamma v_{\lambda_2}^{}(p_2)/(\sqrt{2}M_0)$, where $\Gamma= \gamma_5$ and
$-\hat{\slashed{\epsilon}}(J_z) + \hat{\epsilon}(J_z) \cdot (p_1-p_2)/(M_0 + 2m)$ 
for pseudoscalar and vector mesons, respectively,\footnote{The Lorentz invariant properties with the BT construction discussed here would in general apply to 
other types of the wave function vertices as well, e.g., the axial vector coupling for the pseudoscalar meson vertex in the analysis of axial anomaly.}
and $M^2_0=\sum_i ({\bf k}_{i\perp}^2+m^2_i)/x_i$.
The polarization vectors 
$\hat{\epsilon}^\mu(J_z)$ 
of the vector meson are given by
$\hat{\epsilon}^\mu(\pm1) = \bm{(} 0, 2 \bm{\epsilon}_\perp(\pm1) \cdot {\bf P}_\perp/P^+, \bm{\epsilon}_\perp(\pm1)\bm{)}$ 
with $\bm{\epsilon}_\perp(\pm 1) = \mp \left( 1, \pm i \right)/\sqrt{2}$ for transverse polarizations and
$\hat{\epsilon}^\mu(0) = \bm{(} P^+, {( {\bf P}^2_\perp-M^2_0) }/{P^+}, {\bf P}_\perp \bm{)} / M_0$ 
for longitudinal 
polarization~\cite{Jaus91,Jaus90}.
One of the important characteristics of our LFQM in contrast to other covariant field theoretic computations in  LFD~\cite{DF97,DFPS03,BCJ02} 
is to use $M_0$ other than the physical mass $M$ in defining $\mathcal{R}^{JJ_z}_{\lambda_1\lambda_2}$ and 
$\hat{\epsilon}^\mu(0)$ 
as well.
Because of this property imposed by the on-shellness of the constituents, which is consistent with the BT construction, 
the SO wave functions satisfy the unitary condition,
$\sum_{\lambda_1,\lambda_2} \mathcal{R}^{JJ_z\dagger}_{\lambda_1\lambda_2}\mathcal{R}^{JJ_z}_{\lambda_1\lambda_2}= 1$,
independent of model parameters.
Furthermore, the longitudinal polarization vector satisfies 
$P \cdot \hat{\epsilon}(0)=0$
only when $P=p_1+p_2$ or
equivalently $P^2=M^2_0$, which we call the self-consistency condition.
We should note that the LF energy conservation ($P^-=p^-_1 + p^-_2$)  in addition to the LF three-momentum conservation at the meson-quark vertex is required 
for the calculations of the physical observables using the matrix element with the one-body current to be consistent with the BT construction~\cite{BT53,KP91} up to 
the level of computation presented in this work as the 
meson state is constructed by the noninteracting $Q\bar{Q}$ representations. The interaction between quark and antiquark is implemented in the radial wave function 
through the mass spectroscopic analysis as discussed below.
This condition will be shown to be important in the complete covariant analysis of the meson decay constants in the LFQM.

The interactions between quark and antiquark are included in the mass operator~\cite{BT53,KP91} to compute the mass eigenvalue
of the meson state.
In our LFQM, we treat the radial wave function as a trial function for the variational principle to the QCD-motivated effective Hamiltonian 
$H_{Q\bar{Q}}$, i.e., $H_{Q\bar{Q}}\ket{\Psi}=(M_0 + V_{Q\bar{Q}})\ket{\Psi}= M\ket{\Psi}$, so that the mass eigenvalue  
is obtained from the interaction potential $V_{Q\bar{Q}}$ in addition to the relativistic free energies of quark and antiquark.
The detailed mass spectroscopic analysis can be found in Refs.~\cite{CJLR15,ACJO22}.
For the radial wave function of the $1S$ state meson, we use the Gaussian wave function 
$\phi (x, \mathbf{k}_\bot) =  \sqrt{\partial k_z/\partial x}\ {\hat\phi} (\mathbf{k})$ as a trial wave function, where
${\hat\phi} (\mathbf{k})= (4\pi^{3/4}/ \beta^{3/2}) \exp(-{\bf k}^2/ 2\beta^2)$ and $\beta$ is the variational parameter fixed by mass 
spectroscopic analysis. 
It should be mentioned, however, that our observation and discussion about the independence of the model predictions with respect to the 
components of the current, the polarization vectors, and the reference frames is completely irrelevant to any specific form of the radial 
wave function as far as the Jacobian factor $\sqrt{\partial k_z/\partial x}$, which is 
crucial for the Lorentz invariance of the LFQM, is properly included. 

\mytitle{Decay constants}
The decay constants, $f_{\rm P}$ for the pseudoscalar (P) meson, $f_{\rm V}$ and $f_{\rm V}^{T}$ for the longitudinally and 
transversely polarized vector (V) mesons, with their corresponding one-body currents are defined as
\begin{eqnarray}\label{eq:2}
	\bra{0} \bar{q} \gamma^\mu \gamma_5 q \ket{{\rm P}(P)} &=& i f_{\rm P} P^\mu, \nonumber\\
	\bra{0} \bar{q} \gamma^\mu q \ket{{\rm V}(P,J_z)} &=& f_{\rm V} M \epsilon^\mu(J_z),\\
	\bra{0} \bar{q} \sigma^{\mu\nu} q \ket{{\rm V}(P,J_z)} &=& if_{\rm V}^T 
	\left[\epsilon^\mu(J_z) P^\nu -\epsilon^\nu(J_z) P^\mu \right], \nonumber
\end{eqnarray}
where $P^\mu$ and $M$ are the meson momentum and mass, respectively, 
and $\sigma^{\mu\nu}=i[\gamma^\mu,\gamma^\nu]/2$.

In principle, 
the Lorentz structures in the right-hand side of Eq.~(\ref{eq:2}) should 
be independent of the internal momentum of the quark-antiquark system. 
For instance, the longitudinal polarization vector of the vector meson defined in the right-hand side of Eq.~(\ref{eq:2}) 
should be used with the physical mass $M$, i.e.,
${\epsilon}^\mu(0) = \bm{(} P^+, {( {\bf P}^2_\perp-M^2) }/{P^+}, {\bf P}_\perp \bm{)} / M$.  
Typically, one can obtain the decay constants using some particular choice of
the currents and polarizations to preserve the Lorentz structures as given in the right-hand side of Eq.~(\ref{eq:2})~\cite{Jaus91,CJ07}, 
(i) $f_{\rm P}$ from $\gamma^{(+, \perp)}\gamma_5$,
(ii) $f_{\rm V}$ from $\gamma^{+}$ and $\epsilon(0)$, and
(iii) $f^T_{\rm V}$ from $\sigma^{\perp +}$ and $\epsilon(+1)$ as one can see from Eq.~(\ref{eq:2}). 
Those results of $(f_{\rm P}, f_{\rm V}, f^T_{\rm V})$ obtained from (i)-(iii) have already been provided as the standard 
LFQM results~\cite{CJ07} (see Eqs. (18)-(20) in Ref.~\cite{CJ07}), rewriting 
the decay constants $\mathcal{F} = \{ f_{\rm P}, f_{\rm V}, f^{T}_{\rm V}\}$ as
\begin{eqnarray}\label{eq:3}
\mathcal{F} &=& \sqrt{N_c} \int^1_0 {\rm d}x \int \frac{{\rm d}^2 \mathbf{k}_\bot}{16\pi^3}\  \phi(x,\mathbf{k}_\perp)  
\nonumber\\  &&  \mbox{} \times
\frac{1}{\cal P} \sum_{\lambda_1, \lambda_2} \mathcal{R}_{\lambda_1 \lambda_2}^{JJ_z} 
\left[\frac{\bar{v}_{\lambda_2}(p_2)}{\sqrt{x_2}} {\cal G}  \frac{u_{\lambda_1}(p_1)}{\sqrt{x_1}}\right],
\end{eqnarray}
where $N_c=3$ is the number of color and the current operators ${\cal G} =\{\gamma^\mu\gamma_5,\gamma^\mu,\sigma^{\mu\nu}\}$ 
pair with the corresponding Lorentz structures 
${\cal P}=\{P^\mu, M \epsilon^\mu(J_z),  i[\epsilon^\mu(J_z) P^\nu -\epsilon^\nu(J_z) P^\mu]\}$
defined in the right hand side of Eq.~(\ref{eq:2}).

However, we note here that the Lorentz structures 
${\cal P}=\{P^\mu, M \epsilon^\mu(J_z), i[\epsilon^\mu(J_z) P^\nu -\epsilon^\nu(J_z) P^\mu]\}$ 
in Eq.~(\ref{eq:3}) for the particular choices of the currents and polarizations taken in (i)-(iii) apparently satisfy the self-consistency condition, $P=p_1+p_2$  in ${\cal P}$,
 due to the momentum conservation for the + and $\perp$ components. Such manifest realization of 
 the self-consistency condition cannot be attained for the choices beyond (i)-(iii) taken in the computation. 
 Nevertheless, we realize that the identical self-consistency condition can still be verified by linking the computation of the same physical 
 observables between the manifestly covariant Bethe-Salpeter (BS) model and the standard LFQM as shown in Refs.~\cite{CJ13,Choi21}. 
 Using different components of the currents and polarization vectors such as 
 $f_{\rm P}$ from $\gamma^{-}\gamma_5$~\cite{Choi21} and $f_{\rm V}$ from $\gamma^{\perp}$ and 
$\epsilon(+1)$~\cite{CJ13}, we find in this work that the same self-consistency condition, $P=p_1+p_2$ in ${\cal P}$,
 is applicable to all the Lorentz structures $\mathcal{P}$ in Eq.~(\ref{eq:3}) to attain the complete covariance of the decay constants for 
all possible combinations of currents and polarization vectors including the ones not discussed 
in Refs.~\cite{Jaus91,CJ07,CJ13,Choi21}. As mentioned in the introduction, this self-consistency condition for the one-body current matrix element computation 
is consistent with the BT construction up to that level of computation in which the meson state is constructed by the noninteracting $Q\bar{Q}$ representations while the interaction is included in the mass operator $M:= M_0 + V_{Q\bar{Q}}$. 
\begin{table}
	\begin{ruledtabular}
		\renewcommand{\arraystretch}{1.2}
		\caption{ The operators $\mathcal{O}_{\rm BS}$ defined in Eq.~(\ref{eq:4b}). Note that $\mathcal{O}_{\rm BS}$ turns into $\mathcal{O}_{\rm LFQM}$ 
		if $M\to M_0$ is made, which are exactly the same as those defined in Eq.~(\ref{eq:4})  for the standard LFQM.
		}
		\label{OBS}
		\begin{tabular}{ccc|c|c}
		${\cal F}$	 & ${\cal G}$ & \hspace{-0.2cm} $\epsilon(J_z)$ \hspace{-0.2cm} & $\mathcal{O}_{\rm BS}$  & $\mathcal{O}_{\rm LFQM}$\\  \hline 
			\multirow{3}{*}{$f_{\rm P}$}  & $\gamma^{(+,\perp)}\gamma_5$ & & $2m$ & $2m$\\
			   & \multirow{2}{*}{$\gamma^{-}\gamma_5$}     &  
			   & \multirow{2}{*}{$2m\frac{M^2_0 + \mathbf{P}_\perp^2}{M^2 + {\bf P}^2_\perp} $}   
			   & \multirow{2}{*}{$2m$}  \\ 
			    &&&&\\ \hline 
		    \multirow{6}{*}{$f_{\rm V}$}  & \multirow{2}{*}{$\gamma^{(+,\perp)}$} 
		    & \multirow{2}{*}{$\epsilon(0)$} &  \multirow{2}{*}{$\frac{(M^2_0 +M^2)[m + 2x(1-x)M]}{M(M+2m)}$} 
		    & \multirow{2}{*}{$2m + \frac{4\mathbf{k}_\bot^2}{{\mathcal D}_0}$} \\
		    &&&&\\
			   & \multirow{2}{*}{$\gamma^-$}   
			   & \multirow{2}{*}{$\epsilon(0)$} 
			   & \multirow{2}{*}{$\frac{\hat\epsilon^-(0)M_0(M^2_0 +M^2)[m + 2x(1-x)M]}{\epsilon^-(0)M^2(M+2m)}$} 
                 & \multirow{2}{*}{$2m + \frac{4\mathbf{k}_\bot^2}{{\mathcal D}_0}$} \\
                &&&& \\ \cline{2-5}
			    & \multirow{2}{*}{$\gamma^{(\perp,-)}$} 
			   & \multirow{2}{*}{$\epsilon(+1)$} 
			   & \multirow{2}{*}{$\frac{1}{M}\left(M^2_0 - \frac{2M\mathbf{k}_\perp^2}{M+2m}\right)$} 
                & \multirow{2}{*}{$M_0 - \frac{2\mathbf{k}_\bot^2}{{\mathcal D}_0}$}  \\ 
                &&&&  \\ \hline 
            \multirow{6}{*}{$f^{T}_{\rm V}$} & \multirow{2}{*}{$\sigma^{\perp +}$} & \multirow{2}{*}{$\epsilon(+1)$} & 
            \multirow{2}{*}{$2m + \frac{2\mathbf{k}_\bot^2}{M+2m}$}  &  \multirow{2}{*}{$2m + \frac{2\mathbf{k}_\bot^2}{{\mathcal D}_0}$} \\  
            &&&& \\
            & \multirow{2}{*}{$\sigma^{\perp -}$} & \multirow{2}{*}{$\epsilon(+1)$} 
            & \multirow{2}{*}{$2\frac{M^2_0}{M^2}\left(m + \frac{\mathbf{k}_\bot^2}{M+2m}\right)$} 
            & \multirow{2}{*}{$2m + \frac{2\mathbf{k}_\bot^2}{{\mathcal D}_0}$}\\ 
            &&&& \\ \cline{2-5}
            & \multirow{2}{*}{$\sigma^{+ -}$} & \multirow{2}{*}{$\epsilon(0)$} 
            & \multirow{2}{*}{ $\frac{M^2_0+M^2}{2M^2}\left(\frac{2mM + M^2_0}{M+2m} - \frac{4\mathbf{k}_\perp^2}{M+2m}\right)$ }
            & \multirow{2}{*}{$M_0 - \frac{4\mathbf{k}_\perp^2}{\mathcal{D}_0}$}\\ 
            &&&& \\ 
		\end{tabular}
		\renewcommand{\arraystretch}{1}
	\end{ruledtabular}
\end{table}

\begin{table*}[t]
	\begin{ruledtabular}
		\renewcommand{\arraystretch}{1.2}
		\caption{ The operators $\mathcal{O}_{\rm LFQM}$ and the helicity contributions $H_{\lambda_1\lambda_2}$ to $\mathcal{O}_{\rm LFQM}$ defined in 
		Eq.~(\ref{eq:4}) for all possible components of the current ${\cal G}$ and the polarization vectors $\epsilon(J_z)$, 
		where $x_1=x, x_2=1-x$, and ${\cal D}_0=M_0 +2m$. 
		}
		\label{coeff}
		\begin{tabular}{ccc|cccc|c}
		${\cal F}$	 & ${\cal G}$ & \hspace{0.2cm} $\epsilon(J_z)$ \hspace{0.2cm} & $H_{\uparrow\uparrow}$ & 
		$H_{\uparrow\downarrow}$ & $H_{\downarrow\uparrow}$ 
                     & $H_{\downarrow\downarrow}$  
                     & $\mathcal{O}_{\rm LFQM}$\\  \hline 
			\multirow{3}{*}{$f_{\rm P}$}  & $\gamma^{(+,\perp)}\gamma_5$ & & $0$ & $m$ & $m$ & $0$ & $2m$\\
			   & \multirow{2}{*}{$\gamma^{-}\gamma_5$}     &  
			   & \multirow{2}{*}{$\frac{2m\mathbf{k}_\perp^2}{x_1 x_2(M_0^2 + {\bf P}^2_\perp)} $}   
			   & \multirow{2}{*}{$m - \frac{2m\mathbf{k}_\perp^2}{x_1x_2(M_0^2 + {\bf P}^2_\perp)} $}   
			   & \multirow{2}{*}{$m - \frac{2m\mathbf{k}_\perp^2}{x_1 x_2(M_0^2 + {\bf P}^2_\perp)} $} 
			   & \multirow{2}{*}{$\frac{2m\mathbf{k}_\perp^2}{x_1x_2(M_0^2 + {\bf P}^2_\perp)}$}  
			   & \multirow{2}{*}{$2m$}  \\ 
			    &&&&&&& \\ \hline 
		    \multirow{6}{*}{$f_{\rm V}$}  & \multirow{2}{*}{$\gamma^{(+,\perp)}$} 
		    & \multirow{2}{*}{$\epsilon(0)$} &  \multirow{2}{*}{$0$} 
		    & \multirow{2}{*}{$m + \frac{2\mathbf{k}_\bot^2}{{\mathcal D}_0}$} 
		    & \multirow{2}{*}{$m + \frac{2\mathbf{k}_\bot^2}{{\mathcal D}_0}$} 
		    & \multirow{2}{*}{$0$} 
		    & \multirow{2}{*}{$2m + \frac{4\mathbf{k}_\bot^2}{{\mathcal D}_0}$} \\
		    &&&&&&&\\
			   & \multirow{2}{*}{$\gamma^-$}   
			   & \multirow{2}{*}{$\epsilon(0)$} 
			   & \multirow{2}{*}{0} 
			   & \multirow{2}{*}{$m + \frac{2\mathbf{k}_\bot^2}{{\mathcal D}_0}$} 
                 & \multirow{2}{*}{$m + \frac{2\mathbf{k}_\bot^2}{{\mathcal D}_0} $} 
                 & \multirow{2}{*}{0}
                 & \multirow{2}{*}{$2m + \frac{4\mathbf{k}_\bot^2}{{\mathcal D}_0}$} \\
                &&&&&&& \\ \cline{2-8}
			    & \multirow{2}{*}{$\gamma^{(\perp,-)}$} 
			   & \multirow{2}{*}{$\epsilon(+1)$} 
			   & \multirow{2}{*}{$M_0 - \frac{(M_0 + m)\mathbf{k}_\perp^2}{x_1 x_2 M_0 \mathcal{D}_0}$} 
			   & \multirow{2}{*}{$\frac{x_1 (x_1 M_0 + m)\mathbf{k}^2_\perp}{{x_1 x_2 M_0\mathcal D}_0}$} 
                & \multirow{2}{*}{$\frac{x_2(x_2 M_0 +  m)\mathbf{k}^2_\perp}{x_1 x_2 M_0 {\mathcal D}_0}$}  
                & \multirow{2}{*}{$0$} 
                & \multirow{2}{*}{$M_0 - \frac{2\mathbf{k}_\bot^2}{{\mathcal D}_0}$}  \\ 
                &&&&&&&  \\ \hline 
            \multirow{6}{*}{$f^{T}_{\rm V}$} & \multirow{2}{*}{$\sigma^{\perp +}$} & \multirow{2}{*}{$\epsilon(+1)$} & 
            \multirow{2}{*}{$2m + \frac{2\mathbf{k}_\bot^2}{{\mathcal D}_0}$}  & \multirow{2}{*}{0} & \multirow{2}{*}{0} 
            & \multirow{2}{*}{0} &  \multirow{2}{*}{$2m + \frac{2\mathbf{k}_\bot^2}{{\mathcal D}_0}$} \\  
            &&&&&&& \\
            & \multirow{2}{*}{$\sigma^{\perp -}$} & \multirow{2}{*}{$\epsilon(+1)$} 
            & \multirow{2}{*}{$2m -\frac{2m(m+M_0)\mathbf{k}_\perp^2}{x_1 x_2 M^2_0\mathcal{D}_0}$} 
            & \multirow{2}{*}{$\frac{2m(m+x_1 M_0)\mathbf{k}_\perp^2}{x_1 x_2 M^2_0\mathcal{D}_0}$} 
            & \multirow{2}{*}{$\frac{2m(m+ x_2 M_0)\mathbf{k}_\perp^2}{x_1 x_2 M^2_0\mathcal{D}_0}$} 
            & \multirow{2}{*}{$\frac{2\mathbf{k}_\perp^4}{x_1 x_2 M^2_0\mathcal{D}_0}$} 
            & \multirow{2}{*}{$2m + \frac{2\mathbf{k}_\bot^2}{{\mathcal D}_0}$}\\ 
            &&&&&&& \\ \cline{2-8}
            & \multirow{2}{*}{$\sigma^{+ -}$} & \multirow{2}{*}{$\epsilon(0)$} 
            & \multirow{2}{*}{ $\frac{\mathbf{k}_\perp^2}{2 x_1x_2\mathcal{D}_0} - \frac{2\mathbf{k}_\perp^2}{\mathcal{D}_0}$ }
            & \multirow{2}{*}{$\frac{M_0}{2} - \frac{\mathbf{k}_\perp^2}{2 x_1 x_2\mathcal{D}_0} $}
            & \multirow{2}{*}{$\frac{M_0}{2} - \frac{\mathbf{k}_\perp^2}{2 x_1 x_2\mathcal{D}_0} $}
            & \multirow{2}{*}{ $\frac{\mathbf{k}_\perp^2}{2x_1 x_2\mathcal{D}_0} - \frac{2\mathbf{k}_\perp^2}{\mathcal{D}_0}$ }
            & \multirow{2}{*}{$M_0 - \frac{4\mathbf{k}_\perp^2}{\mathcal{D}_0}$}\\ 
            &&&&&&& \\ 
		\end{tabular}
		\renewcommand{\arraystretch}{1}
	\end{ruledtabular}
\end{table*}

\mytitle{Link between the BS model and the LFQM}
For a full demonstration of the validity of the identical self-consistency condition, $P=p_1+p_2$ or 
$M \to M_0$ in ${\cal P}$, engaging any combination of current component and of 
polarization vector in Eq.(\ref{eq:3}), we briefly discuss the link between the manifestly covariant BS model and the standard LFQM. In the manifestly covariant BS model~\cite{CJ13,Choi21}, 
the generic form of the matrix element  for the decay amplitude $A_{\rm BS}\equiv \langle 0|{\bar q}{\cal G} q|V(P,J_z)\rangle$  
in the one-loop approximation is given by
\begin{eqnarray}\label{eq:4a}
A_{\rm BS}  &=& N_c\int\frac{{\rm d}^4 p_2}{(2\pi)^4}\frac{H_V S_{\rm BS}}{(p^2_1 - m^2 + i\epsilon)(p^2_2 - m^2 + i\epsilon)},
\nonumber\\
&=& N_c\int^1_0 \frac{{\rm d}x}{(1-x)}\int \frac{{\rm d}^2{\bf k}_\perp}{16\pi^3}\chi(x,{\bf k}_\perp) [S_{\rm BS}]_{\rm on},
\end{eqnarray}
where the trace term $S_{\rm BS}={\rm Tr}[{\cal G}(\slashed{p}_{1} + m)\Gamma(-\slashed{p}_{2} + m)]$ in the first line becomes $[S_{\rm BS}]_{\rm on}$ in the second line 
after the light-front energy integration $p_2^-$ picking up the on-mass shell pole
$p_2^2=m^2$ and the resulted light-front BS vertex function $\chi(x,{\bf k}_\perp)$ after the pole integration is given by $\chi(x,{\bf k}_\perp)=g/[x(M^2-M^2_0)]$.
We note that the manifestly covariant meson vertex 
 $\Gamma_{\rm V}= \slashed{\epsilon}(J_z) - (p_1 - p_2)\cdot\epsilon(J_z)/(M + 2m)$ 
carries the longitudinal polarization $\epsilon^\mu(0)$ including the physical meson mass $M$ 
in contrast to the standard LFQM where $\hat\epsilon^\mu(0)$ is used for the spin-orbit wave function.
While we take here a constant $Q{\bar Q}$ bound-state vertex function, 
i.e., $H_V=g$, for simplicity, we should note that the usual multipole ansatz~\cite{CJ13} for the $Q\bar{Q}$ bound-state vertex function such as 
$H_V=g/(p^2 - \Lambda^2 + i\epsilon)^n$ with the parameter
$\Lambda$ only alters the form of $\chi(x,{\bf k}_\perp)$ but not the generic form of Eq.~(\ref{eq:4a}). 
Comparing the computation between the covariant BS model and the standard LFQM, we find that the link, 
i.e., $\sqrt{2N_c}\chi(x, {\bf k}_\perp)/(1-x)\to\phi(x,{\bf k}_\perp)/\sqrt{m^2 + {\bf k}^2_\perp}$ and $M\to M_0$, 
applies to all possible components of the currents and polarization vectors as it has already been found for the case of $f_{\rm P}$ obtained 
from ${\cal G}=(\gamma^+,\gamma^-)\gamma_5$~\cite{Choi21} and $f_{\rm V}$ obtained from $\cal{G}=(\gamma^+,\gamma^\perp)$
with $\bm{(} \epsilon(0), \epsilon(+) \bm{)}$~\cite{CJ13}, respectively. 
One should note that the possible instantaneous and zero-mode contributions vanish
with the above link 
as shown in Refs.~\cite{CJ13,Choi21}.
The instantaneous contribution with the $\gamma^+$ operator 
appears always proportional to $(M^2 - M^2_0)$ and the zero-mode operator found in the two-point function~\cite{CJ13} is proportional 
to $Z_2 = x (M^2 - M^2_0) + (1-2x)M^2$ for the equal quark mass case. These contributions vanish under the link $M\to M_0$ 
discussed in Refs.~\cite{CJ13,Choi21}. Note that the term $(1-2x)M^2$ in $Z_2$ vanishes as well after the replacement of $M\to M_0$ 
because it is an odd function of $x$ while other terms in the integrand are even in $x$ 
as shown in Ref.~\cite{CJ13}
and can be seen later also in this work.
For the complete analysis of $(f_{\rm P}, f_{\rm V}, f^T_{\rm V})$ on the validity of the link between the BS model and the standard LFQM
extending the previous works~\cite{CJ13,Choi21}, 
we show the generic form of the decay constants in Eq.~(\ref{eq:4a}) obtained from the on-mass shell quark propagating part
 as 
\begin{equation}\label{eq:4b}
	{\cal F}_{\rm BS} =N_c \int^1_0  \frac{{\rm d}x}{(1-x)}\int \frac{{\rm d}^2 \mathbf{k}_\bot}{8\pi^3}~\chi(x,{\bf k}_\perp)\mathcal{O}_{\rm BS}(x,{\bf k}_\perp),
\end{equation} 
where the operators $\mathcal{O}_{\rm BS}$ are defined by $\mathcal{O}_{\rm BS}=[S_{\rm BS}]_{\rm on}/2{\cal P}$, and
$\mathcal{O}_{\rm BS}=\{\mathcal{O}_{\rm P}, \mathcal{O}_{\rm V}(J_z),\mathcal{O}^{T}_{\rm V}(J_z)\}$
corresponding to $\mathcal{F}_{\rm BS} = \{ f_{\rm P}, f_{\rm V}, f^{T}_{\rm V}\}$ for the equal quark and antiquark mass case
are summarized in Table~\ref{OBS}.

As we shall show later in Eq.~(\ref{eq:4}), the standard LFQM results $\mathcal{F}$ obtained directly from Eq.~(\ref{eq:3}) are indeed exactly 
the same as the ones obtained from  $\mathcal{F}_{\rm BS}$ applying the ``Type II"~\cite{CJ13}
link, i.e., $\sqrt{2N_c}\chi(x, {\bf k}_\perp)/(1-x)\to\phi(x,{\bf k}_\perp)/\sqrt{m^2 + {\bf k}^2_\perp}$ and $M\to M_0$, in Eq.~(\ref{eq:4b}). 
The corresponding operators $\mathcal{O}_{\rm LFQM}$ obtained from replacement of $M\to M_0$ in $\mathcal{O}_{\rm BS}$
are also summarized in Table~\ref{OBS}.
In other words, the same self-consistency condition, $P=p_1+p_2$ or $M \to M_0$ in ${\cal P}$, 
should be applied to all the Lorentz structures $\mathcal{P}$ in Eq.~(\ref{eq:3}) to attain the complete covariance of the decay constants in the standard LFQM
for all possible combinations of currents and polarization vectors including the ones not discussed in Refs.~\cite{Jaus91,CJ07,CJ13,Choi21}.

In the covariant BS model, we also note that some combinations of the current components and polarization vectors~\cite{CJ13,Choi21} 
encounter the LF zero modes and give correct results only if the zero-mode contributions are not missed but taken into account properly. 
One may note from Table~\ref{OBS} that only the operator ${\cal O}_{\rm BS}= 2m$ for $f_{\rm P}$ obtained from $\gamma^{(+,\perp)}\gamma_5$ 
exactly matches with ${\cal O}_{\rm LFQM}$ in the standard LFQM, indicating that all other BS results for the decay constants except that case would
require zero mode contributions to give correct covariant results.

As the zero-mode contribution is locked into a single point of the LF longitudinal momentum in the meson decay process, one of the constituents of the meson 
carries the entire momentum of the meson, and it is important to capture the effect from a pair creation of particles with zero LF longitudinal momenta indicating an intensive interaction with the vacuum. 
The zero modes appeared for some particular combinations of the current and polarization in the BS model
are found to match with the substitution of $M\to M_0$ for those combinations in the standard LFQM. 
The present analysis of the meson decay constant with all possible combinations of the current and polarization confirmed the previous 
interpretation~\cite{CJ13} for the substitution $M\to M_0$ in the standard LFQM with effective degrees of freedom represented 
by the constituent quark and antiquark as providing the view of an effective zero-mode cloud around the quark and antiquark inside the meson.

In a nutshell,  
we show the explicit final formula of the decay constants directly obtained from Eq.~(\ref{eq:3}) for the equal quark and antiquark mass case:
\begin{eqnarray}\label{eq:4}
	{\cal F} = \sqrt{6} \int^1_0 {\rm d}x\int \frac{{\rm d}^2 \mathbf{k}_\bot}{16\pi^3}\  
	\frac{ {\phi}(x, \mathbf{k}_\bot) }{\sqrt{m^2 + \mathbf{k}_\bot^2}} ~\mathcal{O}_{\rm LFQM}(x,{\bf k}_\perp),
\end{eqnarray}
where the operators $\mathcal{O}_{\rm LFQM}=\{\mathcal{O}_{\rm P}, \mathcal{O}_{\rm V}(J_z),\mathcal{O}^{T}_{\rm V}(J_z)\}$
corresponding to $\mathcal{F} = \{ f_{\rm P}, f_{\rm V}, f^{T}_{\rm V}\}$, respectively, are obtained from the sum of each helicity contribution, 
$\mathcal{O}_{\rm LFQM}=\sum_{\lambda_1,\lambda_2} H_{\lambda_1\lambda_2}$.
It should be noted that Eq.~(\ref{eq:4}) is a generalized formula for the previous standard LFQM results~\cite{CJ07} for $(f_{\rm P}, f^{(T)}_{\rm V})$ 
where the substitution $M\to M_0$ is manifest due to the + and $\perp$ momentum conservation. 
Equation~(\ref{eq:4}) is indeed exactly the same as the one obtained from applying the link between the BS model and the
standard LFQM to Eq.~(\ref{eq:4b}).

We summarize our results of $\mathcal{O}_{\rm LFQM}$ and the helicity contributions $H_{\lambda_1\lambda_2}$ to 
$\mathcal{O}_{\rm LFQM}$ for all possible components of the current $\mathcal{G}$ and the polarization vectors $\epsilon(J_z)$
in Table~\ref{coeff}. 
The results of $J_z = -1$ are not explicitly given for $f_{\rm V} $ and $f^{T}_{\rm V}$ as they correspond to those of 
$J_z=+1$ with $H_{\lambda_1\lambda_2}(J_z=-1) = H_{-\lambda_1-\lambda_2}(J_z=+1)$ absorbing the usual 
parity-related phase factor~\cite{JW59,CJ03} within the definition of $H_{\lambda_1\lambda_2}$ as the contribution leading 
to the identical $\mathcal{O}_{\rm LFQM}$ after summing over the helicities. 
To obtain the results, we used the Dirac spinor basis with the chiral representation defined in Refs.~\cite{BPP97,Jaus90}.
The combinations of the current components and polarizations shown in Table~\ref{coeff} are the complete set and
other combinations are not possible to extract the decay constants.
Equation~(\ref{eq:4}) shows that the decay constants are not dependent on the energy of the bound states but on the mass of the constituents. 
This feature reflects the BT construction with the noninteracting $Q\bar{Q}$ representations 
including the interaction only in the mass operator $M:= M_0 + V_{Q\bar{Q}}$ and appears essential for the Lorentz-invariant quark phenomenology of decay constants in the LFQM.

\mytitle{Observation and Discussion}
The results shown in Eq.~(\ref{eq:4}) and Table~\ref{coeff} exhibit the Lorentz invariance of the physical observables 
represented by the decay constant ${\cal F}$, although each helicity contribution $H_{\lambda_1\lambda_2}$ obtained 
in our LFQM apparently depends on (a) the current components $(\mu=\pm,\perp)$, (b) the polarization vectors 
$\epsilon^\mu(J_z)$, and (c) the transverse momentum ${\bf P}_\perp$ of the meson. 
We find that the decay constants ${\cal F}$ resulted by integrating the sum of all helicity contributions, 
$\mathcal{O}_{\rm LFQM}=\sum_{\lambda_1,\lambda_2} H_{\lambda_1\lambda_2}$, with the radial wave function   
$\phi(x, \mathbf{k}_\bot)$ turn out to be completely independent of (a), (b), and (c) and yield unique predictions of our LFQM.

\begin{figure}[t]
	\centering
	\includegraphics[width=1\columnwidth]{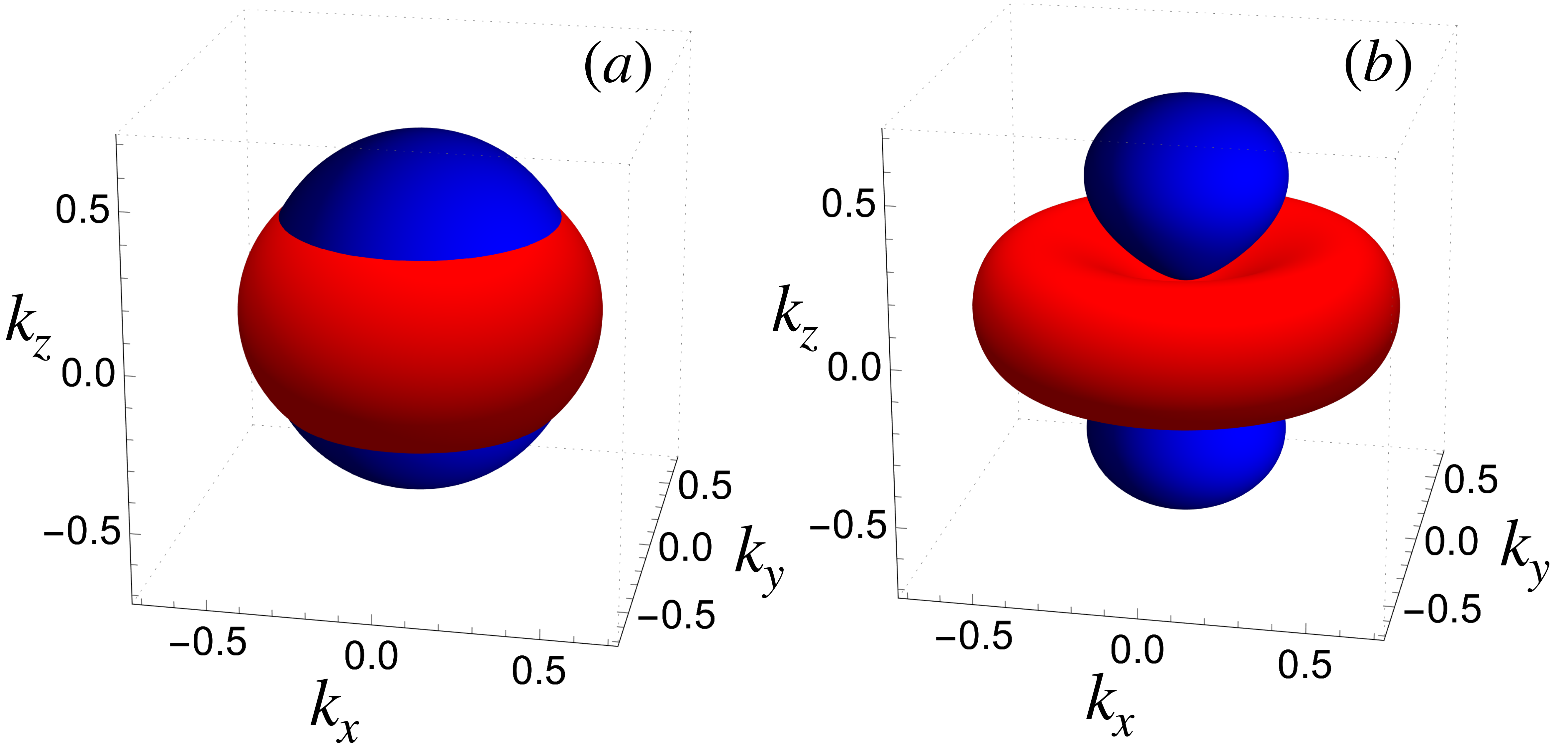}
	\caption{\label{fig1} 
	The 3D plots of the wave functions (a) $\psi^{(J_z)}_{\rho}({\bf k})$ for the $\rho$ meson 
	and (b) ${\tilde\psi}_{\rho}({\bf k})=\psi^{(0)}_{\rho}-\psi^{(+1)}_{\rho}$ defined 
	by $f_\rho (J_z) = \int {\rm d}^3 \mathbf{k}\; \psi^{(J_z)}_{\rho} ({\bf k})$, where $\psi^{(0)}_{\rho}$ (red) 
	and $\psi^{(+1)}_{\rho}$ (blue).}
\end{figure}

For the quantitative estimation of decay constants, we exemplify the $(\pi, \rho)$ mesons since they are good 
examples of the relativistic $Q{\bar Q}$ bound states.
The model parameters are chosen as $(m, \beta)=(0.25, 0.3194)$~GeV following Refs.~\cite{CJ97,CJ99a,CJ07}. 
This parameter set gives $f_{\pi}=131$~MeV, $f_{\rho}=215$~MeV, and $f^T_{\rho}=173$ MeV~\cite{CJ07}, 
which are in a good agreement with the experimental data, $f^{\rm Expt.}_{\pi}=130.3\pm 0.3$~MeV and 
$f^{\rm Expt.}_{\rho}=210\pm 4$~MeV~\cite{PDG22}.
However, what we would like to stress here is the uniqueness of the model predictions on the physical observables
beyond just a good agreement with the data. 
Namely, the decay constant predicted by our LFQM is identical regardless of the aforementioned (a), (b), and (c).
In particular, it is remarkable to see from Table~\ref{coeff} that our analytic forms of the decay constants completely 
satisfy the SU(6) symmetry relation~\cite{Leutwyler74}, $f_{\rm P} + f_{\rm V}(J_z) = 2 f^T_{\rm V}(J_z)$,
for each polarization vector $\epsilon(J_z)$ of the vector meson regardless of the components of the currents used
in the calculation.
Although the analytic forms of $f^{(T)}_{\rm V}(J_z)$ do not look same for different $J_z$, they are in fact the same. 
This can be shown explicitly by converting Eq.~(\ref{eq:4}) into the integral form of the ordinary three vector 
${\bf k} = (k_z, {\bf k}_\perp)$ by taking into account the Jacobian of the variable transformation, 
$\{x, {\bf k}_\perp \} \to \{k_z, {\bf k}_\perp \}$, i.e., 
\begin{eqnarray}\label{eq:5}
	{\cal F} &=& \sqrt{6}  \int \frac{{\rm d}^3 \mathbf{k} }{(2\pi)^3 }\  
	\frac{ {\hat\phi}(\mathbf{k}) }{M^{3/2}_0} ~\mathcal{O}_{\rm LFQM}({\bf k}),
\end{eqnarray}
where $M_0= 2\sqrt{m^2 + {\bf k}^2}$ and ${\hat\phi}(\mathbf{k})$ corresponds to ${\phi}(x, \mathbf{k}_\bot)$ 
under the variable change $\{x, {\bf k}_\perp \} \to \{k_z, {\bf k}_\perp \}$.
The difference of the two operators $\tilde{\mathcal{O}}^{(T)}_{\rm V}= \mathcal{O}^{(T)}_{\rm V}(J_z=1)-\mathcal{O}^{(T)}_{\rm V}(J_z=0)$ 
is then obtained as 
\begin{eqnarray}
\tilde{\mathcal{O}}^{(T)}_{\rm V} = \frac{2}{\mathcal{D}_0} ({\bf k}^2_\perp - 2 k^2_z)
\end{eqnarray} 
and the integration of $\tilde{\mathcal{O}}^{(T)}_{\rm V}$ in Eq.~(\ref{eq:5}) vanishes since the integrand except 
the term $({\bf k}^2_\perp - 2 k^2_z)$ is rotationally invariant. 
This proves that $f^{(T)}_{V}(J_z=1)=f^{(T)}_{V}(J_z=0)$.
Defining the integrand $\psi^{(J_z)}_{V} ({\bf k})$ for the computation of $f_V (J_z)$ as 
$f_V (J_z) = \int {\rm d}^3 \mathbf{k}\; \psi^{(J_z)}_{V} ({\bf k})$, we display 3D plots of $\psi^{(J_z)}_{\rho}$ for 
the longitudinally polarized $\rho$ meson with $J_z=(0, +1)$ and their difference 
${\tilde\psi}_{\rho}({\bf k})=\psi^{(0)}_{\rho}-\psi^{(+1)}_{\rho}$ in Fig.~\ref{fig1}. 
As one can see, $\psi^{(0)}_{\rho}$ and $\psi^{(+1)}_{\rho}$ show the oblate and prolate ellipsoids, respectively, 
and their difference ${\tilde\psi}_{\rho}({\bf k})\propto (2 k^2_z - {\bf k}^2_\perp)$ reveals the $d$-wave orbital
corresponding to the spherical harmonic function $Y_{20} \propto (3z^2 -r^2)$.
For the transversely polarized $\rho$ meson, the shapes of  $\psi^{T(J_z)}_{\rho}({\bf k})$ are very similar to those of  
$\psi^{(J_z)}_{\rho}({\bf k})$.  
The shape of  $\psi_{\pi} ({\bf k})$, on the other hand, shows the complete spherical symmetry.

\begin{figure}[t]
	\centering
	\includegraphics[width=0.9\columnwidth]{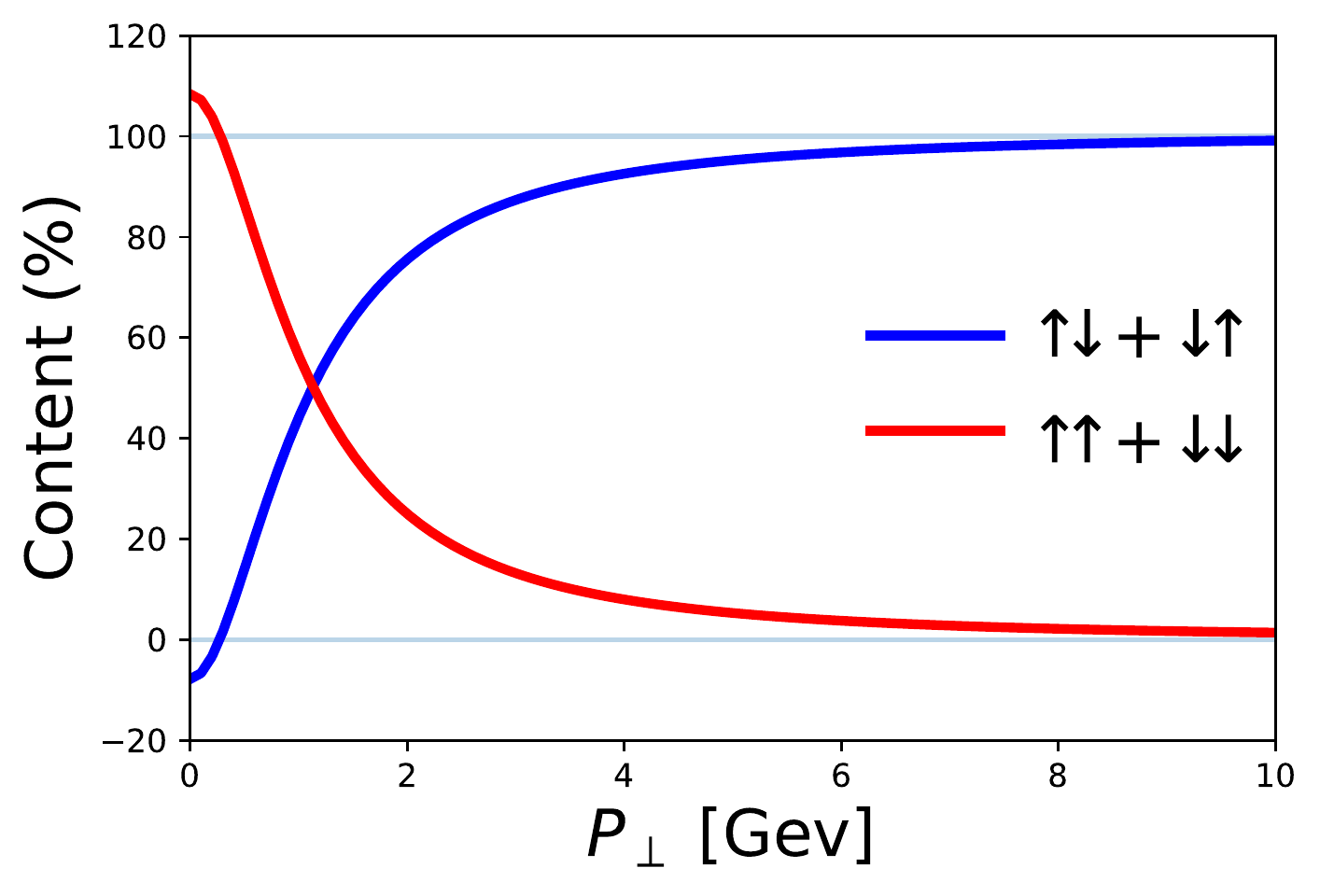}
	\caption{\label{pion} 
	The relative helicity contributions to $f_\pi$ as a function of ${\bf P}_\perp$ calculated with the minus current. The blue and red lines represent the ordinary helicity
	($\uparrow\downarrow,\downarrow\uparrow$) 
and the higher helicity ($\uparrow\uparrow,\downarrow\downarrow$) contributions, respectively. 
	The sum is always the same regardless of ${\bf P}_\perp$.}
\end{figure}

The ${\bf P}_\perp$-independence of our results deserves the remarks below.
As one can see from Table~\ref{coeff}, not only the final operators $\mathcal{O}_{\rm LFQM}$ but also each helicity contributions to  
$\mathcal{O}_{\rm LFQM}$ for the cases of $f^{(T)}_{\rm V}$ and $f_{\rm P}$ with $J^\mu=(J^+, {\bf J}_\perp)$ are obtained to be independent of 
${\bf P}_\perp$ for the equal quark mass case. 
For the case of $f_{\rm P}$ with the minus current, however, each helicity contribution depends on ${\bf P}_\perp$ while the final 
operator ${\cal O}_{\rm P}$ is independent of ${\bf P}_\perp$.
For the illustration of ${\bf P}_\perp$-independence of the final result in the case of the minus current, we show in Fig.~\ref{pion} 
the relative helicity contributions to $f_\pi$ ($\approx 131$~MeV) as a function of ${\bf P}_\perp$. 
The blue and red lines represent the ordinary helicity ($\uparrow\downarrow,\downarrow\uparrow$) 
and the higher helicity ($\uparrow\uparrow,\downarrow\downarrow$) contributions, respectively.
The higher helicity contributions are apparently important for the low and intermediate ${\bf P}_\perp$ regions although 
only the ordinary helicity contribution survives for the ${\bf P}_\perp \to \infty$ limit as in the case of plus and transverse components.

Although $\mathcal{O}_{\rm P}^- = \mathcal{O}_{\rm P}^{(+,\perp)}$ attained for the equal mass case looks rather trivial, 
we note that $\mathcal{O}_{\rm P}^-$ has in fact more complicated structure in the unequal-mass case~\cite{ACJO22}. 
For $\tilde{\mathcal{O}}_{\rm P} \equiv \mathcal{O}_{\rm P}^- - \mathcal{O}_{\rm P}^+$, we find 
\begin{eqnarray}
\label{OpDiff}
    \tilde{\mathcal{O}}_{\rm P} = \frac{ 4(m_1-m_2) M_0 }{(\mathbf{P}_\perp^2 + M_0^2)}k_z 
\end{eqnarray}
for the unequal-mass case. 
While the result of $f_{\rm P}^- = f_{\rm P}^+$ is rather trivial in the equal mass case due to the factor of $m_1-m_2$ in 
Eq.~(\ref{OpDiff}), it is highly nontrivial that this equality $f_{\rm P}^- = f_{\rm P}^+$ prevails even in the unequal mass case. 
The quantity $\tilde{\mathcal{O}}_{\rm P}$ contains the odd-power of $k_z$ as one may intuitively anticipate its appearance from 
$p^- - p^+ = -2 p^3$. 
For the case that $m_1 \neq m_2$, we have $k_z = (x-1/2)M_0+ (m_2^2 - m_1^2)/(2M_0)$ with $M_0 = \sqrt{m_1^2+{\bf k}^2}+ \sqrt{m_2^2+{\bf k}^2}$,
and the corresponding Jacobian $\sqrt{\partial k_z/\partial x}$ included in the radial wave function $\phi (x, \mathbf{k}_\bot)$ recovers 
the same spherically symmetric factor ${\hat\phi}({\bf k})M^{-3/2}_0$ in the integrand of Eq.~(\ref{eq:5}). 
The result of $f_{\rm P}^- = f_{\rm P}^+$ in the unequal mass case is thus due to the symmetry under 
$k_z \leftrightarrow -k_z$ for all other terms beside $\tilde{\mathcal{O}}_{\rm P}$ in the integration. 
Similar behavior is also observed for the case of $f_{\rm V}^-(0) = f_{\rm V}^+(0)$.
These results indicate that one should make sure that the rotational symmetry is not explicitly broken in the wave function level, if one constructs the radial wave function by assuming the separation of the longitudinal and transverse components~\cite{LLCLV21}, e.g., $\phi(x,\mathbf{k}_\perp) = \chi(x) \psi(\mathbf{k}_\perp)$.

\mytitle{Conclusion}
To assert the complete covariance of the decay constants defined by the matrix elements of one-body currents, 
it should be shown that they are completely independent of the current components $(\mu=\pm, \perp)$
and the polarization vectors $(J_z=\pm1, 0)$. 
In this work, for the first time in the standard LFQM, we show this complete covariance by analyzing all the possible components 
of the currents and polarization vectors in the general LF frame with ${\bf P}_\perp\neq 0$.

From the analysis of the respective one-body current matrix elements in LFQM consistent with the BT construction 
at the level of one-body current computation, we obtained the complete Lorentz-invariant results of the decay constants, 
$(f_{\rm P}, f_{\rm V}, f^T_{\rm V})$. 
We analyzed all possible combinations of the current components and the polarizations in the $\mathbf{P}_\perp \neq 0$ frame 
applying the self-consistency condition, $P=p_1 + p_2$ or equivalently $M\to M_0$. 
This condition reflects effectively the BT construction in the computation of the one-body current matrix elements where 
the meson state is described in the non-interacting $Q{\bar Q}$ basis while the interaction is added to the mass operator 
via $M:= M_0 + V_{Q{\bar Q}}$.

 It is important to realize that the decay constants give identical results for the Fock space saturated to the $Q{\bar Q}$ state. 
 While the equivalence should not be limited in principle by the Fock space truncation, it would deserve further analyses 
 to explore the higher Fock states in practice regarding the issue of the cluster separability for the systems of more than 
 two constituents~\cite{Keister-Polyzou-2012}.
In addition to the frame-independence of the results, the verification of the identical results for the physical observables 
regardless of the current components and the polarizations taken in the computation can be used as an important guideline 
for the inclusion of the higher Fock space. 
It is also worthy to mention that the self-consistency condition for the calculation of the matrix elements with one-body current
has been successfully applied to other higher-twist distribution amplitudes of pseudoscalar mesons and semileptonic and 
rare decays between two pseudoscalar mesons~\cite{CJ14,CJ17,Choi21,Choi21b}.
Further applications of our method to other exclusive processes of mesons are under investigation.

\mytitle{Acknowledgements}
We are grateful to Wayne Polyzou and Meijian Li for fruitful discussions.
A.J.A. was supported by the Young Scientist Training (YST) Program at the Asia Pacific Center for Theoretical Physics (APCTP)
through the Science and Technology Promotion Fund and Lottery Fund of the Korean Government and also by the Korean Local 
Governments -- Gyeongsangbuk-do Province and Pohang City.
The work of H.-M.C. was supported by the National Research Foundation of Korea (NRF) under Grant No. NRF- 2020R1F1A1067990.
The work of C.-R.J. was supported in part by the U.S. Department of Energy (Grant No. DE-FG02-03ER41260). 
The National Energy Research Scientific Computing Center (NERSC) supported by the Office of Science of the U.S. Department of Energy 
under Contract No. DE-AC02-05CH11231 is also acknowledged. 
Y.O. was supported by NRF under Grants No. NRF-2020R1A2C1007597 and No. NRF-2018R1A6A1A06024970 (Basic Science Research Program).
The hospitality of the APCTP Senior Advisory Group is gratefully acknowledged.

\end{document}